\documentstyle[12pt,openbib,psfig]{article}
\newcommand{\f}{\frac}

\newcommand{\La}{\Lambda}

\newcommand{\Om}{\Omega}
\newcommand{\be}{\begin{equation}}
\newcommand{\ee}{\end{equation}}

\newcommand{\g}{\gamma}
\newcommand{\om}{\omega}

\begin{document}
\title{\bf Glimpses of a strange star}
\author{Jishnu Dey $^{1,2}$, Subharthi Ray $^{1,2,3}$, X.-D. Li $^4$, Mira
Dey $^{1,3}$  \\ and Ignazio Bombaci $^{5}$}
\date{\today }
\maketitle
\vskip .5cm
\vskip .5cm 
\noindent (1) Abdus Salam ICTP, Trieste, Italy and
Azad Physics Centre, Maulana Azad College, Calcutta 700013, India;
email: azad@vsnl.com \\
(2) 1/10 Prince Golam Md. Road, Calcutta 700 026,India; email :
deyjm@giascl01.vsnl.net.in\\ 
(3) Dept. of Physics, Presidency College, Calcutta 700 073,
India and Abdus Salam ICTP, Trieste, Italy \\ 
(4) Department of Astronomy and Astronomical and Astrophysical Center 
of East China, Nanjing University, Nanjing 210093, China; email:
lixd@nju.edu.cn\\ 
(5) Dipartimento di Fisica, Universit\'a di
Pisa, via Buonarroti 2, I-56127 and INFN Sezione Pisa, Italy;
email: bombaci@pi.infn.it
\newpage

{\bf There are about 2000 gamma ray burst (GRB) events known to us
with data pouring in at the rate of one per day. While the
afterglows of GRBs in radio, optical and X-ray bands are
successfully explained by the fireball model, a significant
difficulty with the proposed mechanisms for GRBs is that a small amount 
($\le 10^{-6}\,M_{\odot}$) of baryons in the ejecta can be involved.
There are very few models that fulfill this criteria together with
other observational features, among which are the differentially
rotating collapsed object model \cite{kr,spruit} and the
"supernova" model \cite{vs}. These models generally invoke rapidly
rotating neutron stars, and may be subject to uncertainties in the
formation mechanisms and the equations of state of neutron stars.
According to Spruit \cite {spruit}, the problem of making a GRB from 
an X-ray binary is reduced to finding a
plausible way to make the star rotate differentially.
We suggest that a model of strange star (SS) can naturally explain
many of these bursts with not only their low baryon content
\cite{dl}, but the differential rotation which leads to an
enhanced magnetic field that surfaces up and is responsible for
GRBs.}

%

The model of SS that we have suggested for some compact objects
\cite{dbdrs, lbddh, lrddb}, has differential rotation
as a consequence of its stratified structure as a natural phenomenon.
It is based on a stable point in the binding energy as a function of
density, for charge neutral, beta stable strange quark matter at about 5
$\rho_0$ where $\rho_0$ is the normal nuclear matter density. It employs
a quark-quark (qq) potential that has asymptotic freedom and 
confinement--deconfinement mechanism built into it.  
At high surface density of $\sim 5\; \rho_0$ at the radius $R$ of the star, 
the qq - interaction is already small as compared to that in a hadron.  
Obviously the interaction is even smaller at the central density of    
$\sim 15 \;\rho_0$.  
Further we have a density dependent mass for the quarks
which at such high central density ensures that the quarks have nearly
current masses. The denser inner parts of the star are composed of
quarks which are asymptotically free and nearly massless whereas the
surface quarks are relatively more massive and interacting - leading to
a peculiar structure which is different from that of a neutron star.

\begin{figure} [h]
\hbox{\hspace{6em}
\hbox{\psfig{figure=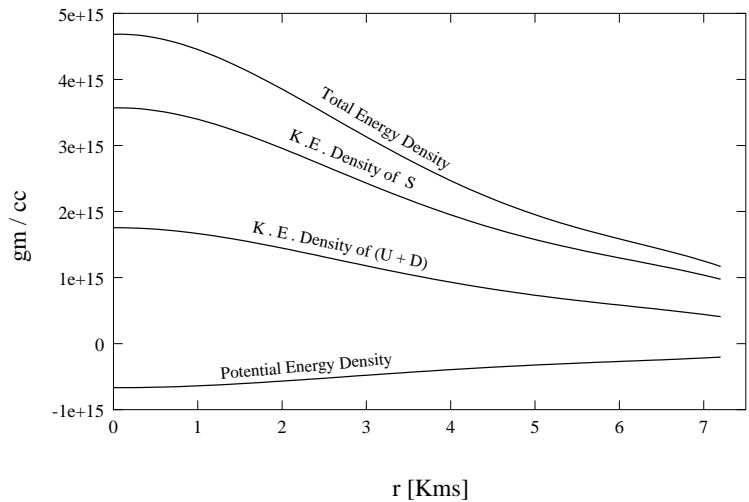,height=6.5cm,angle=-90}}}
\caption{The total energy density, kinetic energy densities (KE) of the
quarks and the potential energy density as functions of the radius inside a
strange star of mass 1.437 $M_\odot$. Recall that for relativistic systems
the KE includes the mass so that the KE(S) dominates.}
\label{fig:kluz1}
\end{figure}

     The energy density as a function of the radius $r$ is shown in Fig.
(\ref{fig:kluz1}). To illustrate the peculiarity of the system we have also
plotted the kinetic energy density (KE) of the quarks. The KE of the u and d
quarks are each roughly half of the total which is less than the KE from the
strange quark. The potential energy is negative and cannot be separated into
parts. The interesting point to see is that the potential energy increases a
little from the surface to the centre but not as much as one would expect,
considering that the number density in the centre is about five times
more than that near the surface. One should recall that the potential energy
is a two-body term and thus is proportional to the square of the number.

\begin{figure} [h]
\hbox{\hspace{6em}
\hbox{\psfig{figure=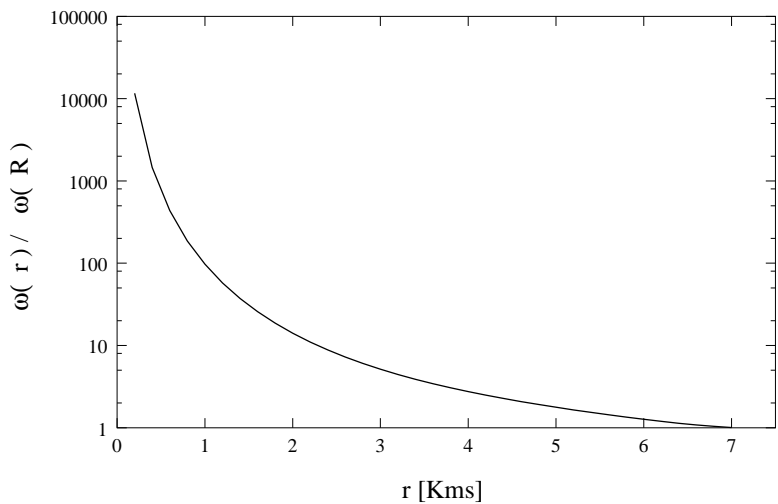,height=6.5cm,angle=-90}}}
\caption{The circular frequency in the equatorial plane at various radii
inside a strange quark star  }
\label{fig:kluz2}
\end{figure}

Using this density variation of we put the surface $r~=~R$ into rotation with
a frequency $\om$(R) about an axis. One can easily see that the central
region on the equatorial plane perpendicular to this axis rotates more than
100 times faster than the outer parts Fig. (\ref{fig:kluz2}) to conserve
angular momentum.  The polar regions will rotate with $\om$(R).  This natural
differential rotation is the required one of the Klu\'zniak and Ruderman
model \cite{kr}.

    According to this model, in a differentially rotating strange star,
the internal poloidal magnetic field ($B_0$) will be wound up into a
toroidal configuration and amplified (to $B_\phi$) as the interior part
of the star rotates faster than the exterior. After $N_{\phi}$
revolutions $B_\phi =2\pi B_0 N_{\phi}$. The field thus amplified forms
a toroid that encloses some strange quark matter. This magnetic toroid
will float up from the deep interior only when a critical field value is
reached that is sufficient to fully overcome the (approximately radial)
stratification in the composition of the strange star.

    The model for strange quark matter that we have proposed is simple and 
is based on 't Hooft's pioneering work on large colour expansion \cite{tH}. 
His idea was to consider the number of colours, $N_c$ in quantum 
chromodynamics to be a parameter of expansion for the field theoretic 
diagrams entering the expressions for variables like the energy of 
quark and gluon fields.
Simple arguments then show that one can obtain a finite theory if one
scales down the quark-gluon or gluon-gluon couplings by a factor of
$N_c^{1/2}$. Then the quark loops are suppressed by a factor of
$\f{1}{N_c}$ compared to planar gluon loops and
non-planar gluon loops are suppressed by $\f{1}{N_c^2}$. As explained by
Witten \cite{wit}, 't Hooft's expansion gives only tree level
interactions between valence quarks for baryons in leading order in the
$\f{1}{N_c}$ - expansion scheme.

    The success of the large colour expansion model has also been
stressed by \cite{wit} and others over the years. In particular, using a
model potential designed to fit the heavier mesons \cite{rich}, as well
as lighter ones like the $\rho, a, f$ - meson \cite{ca,ddt} were
able to fit baryons like the $\Om_-$ and others \cite{ddms} using self
consistent relativistic Hartree-Fock calculations. As already indicated
the potential \cite{rich} has asymptotic freedom and
confinement-deconfinement built into it. This is done very simply and
ingeniously by modifying logarithmic momentum dependence of the running
coupling constant in the potential :
\be
V(q^2) = \f{12 \pi}{27}\f{1}{ln(q^2/\La ^2)}\f{1}{q^2}  \,\, ,
\label{eq:al} \ee to by replacing $\f{1}{q^2}ln(q^2/\La^2)$ by
$\f{1}{q^2}ln(1 + q^2/\La^2)$. For large $q^2$ the original
coupling eq.(\ref{eq:al}) is recovered whereas for large distance
interaction when the momentum transfer $q^2$ is small one gets a
$\f{1}{q^4}$ dependence equivalent to a string-like tension $\La^2
|\vec{r}_1 - \vec{r}_2|$ in the particle coordinates. For dense
systems the $q^2$ is replaced by $q^2 + D^{-2}$ where $D^{-1}$ is
the well known Debye screening factor.

    One further ingredient is very important and this has to do with the
fact that QCD possesses approximate chiral symmetry in the sense
that the current quark masses of u, d, s are small but in the
grounds state this symmetry is broken leading to massive so -
called constituent quarks. At high density chiral symmetry is
believed to be restored and this has been parameterized by us with
a single parameter $\nu$ in a form :
\be
M_i = m_i + 310~~sech(\nu \f{\rho}{\rho_0}). \label{eq:qm}
\ee
where $m_i$ = 4, 7 and 150 for i = u, d and s respectively (all in MeV).  
The suggestion that compact objects like SAX J1808.8-3658, Her X-1,
4U~1820-30 or 4U~1728-34 \cite {dbdrs, lbddh, lrddb} are strange stars 
give us the possibility to fix the chiral symmetry restoration parameter $\nu$ 
(eq. 2) from astronomical data.  
This amounts to constraining microscopic physics of light objects in terms of
some of the densest objects known in the universe.  Although
our model is simple the basis is robust and we believe that the results will
retain their validity even if more refined calculations are done in the
future.

We would like to point out that the strange star candidate SAX
J1808.8-3658 is the fastest rotating X-ray pulsar with surface
rotation frequency $\om(R) \simeq 400$ Hz, shown in the power
spectrum as a very sharp line at that frequency in Wijnands and
van der Klis \cite{wk}. It was suggested that it may appear as an
eclipsing radio pulsar during periods of X-ray quiescence by
Chakraborty and Morgan \cite{cm}. Recently this has been confirmed
when radio signals were found to be present, a day after the X-ray
flux suddenly deviated from exponential decay and began to
decrease rapidly \cite{gsg}, suggesting that LMXB-s are progenitors 
of millisecond radio pulsars
(MSR). This completes the following scenario : the birth of a
strange star may be due to accretion from its binary partner -
leading sometimes to such high rotational frequency that the star
explodes to a GRB. Those that survive due to slower rotation
become LMXB-s like the SAX J1808.8-3658. it may continue to prey
on its partner and become closely related `black widow' MSR which
are evaporating their companions through irradiation as suggested
in \cite{cm}. From the stability of SAX J1808.8-3658 we can safely
assert that only those strange stars rotating faster than a
critical $\om(R)_{crit} > 400$ Hz may acquire the critical
magnetic field and fly off to a GRB mode.

There are several possible channels for strange star formation: type II/Ib
supernovae, accretion-induced collapse of white dwarfs, and conversion from
accreting neutron stars in binary systems \cite{bd2000}. 
The new born strange stars could 
rotate at periods $\le 1$ ms because of rapid rotation of the progenitor
stars due to either contraction or mass accretion.  Furthermore, they are not
subject to the $r-$mode instability \cite{madsen} which slows rapidly
rotating, hot neutron stars to relatively long rotation periods via
gravitational wave radiation. Thus differential rotation may naturally occur
in the interiors of these strange stars as discussed above.

\vskip 0.2cm
The authors SR, MD and JD are grateful to Abdus Salam ICTP,  the
IAEA and the UNESCO for hospitality at Trieste, Italy, and to
Dept. of Science and Technology, Govt. of India. We dedicate this
letter to the memory of Dr. Bhaskar Datta who encouraged this work
through many discussions and was collaborator to one of us (I.B.).
XL was supported by National Natural Science Foundation of China.

\end{document}